\documentclass[%
 aip,
 amsmath,amssymb,
 reprint,%
]{revtex4-1}

\usepackage{graphicx}% Include figure files
\usepackage{dcolumn}% Align table columns on decimal point
\usepackage{bm}% bold math
\usepackage[utf8]{inputenc}
\usepackage[T1]{fontenc}
\usepackage{mathptmx}

\usepackage[version=3]{mhchem} % Formula subscripts using \ce{}
\usepackage{booktabs}
\usepackage{tabularx}
\usepackage{adjustbox}
\usepackage{rotating}
\usepackage{color,soul}
\usepackage{multirow}
\usepackage{relsize}

\usepackage{hyperref}
\hypersetup{
    unicode=true,          % non-Latin characters in Acrobat’s bookmarks
    pdftoolbar=true,        % show Acrobat’s toolbar?
    pdfmenubar=true,        % show Acrobat’s menu?
    pdffitwindow=false,     % window fit to page when opened
    pdfstartview={FitH},    % fits the width of the page to the window
    colorlinks=true,       % false: boxed links; true: colored links
    linkcolor=blue,          % color of internal links
    citecolor=blue,        % color of links to bibliography
    urlcolor=blue,
}

\begin{document}

\preprint{AIP/123-QED}

\title{Self-Interaction Correction in Water--Ion Clusters} 
% Force line breaks with \\

\author{Kamal Wagle}
\email{kamal.wagle@temple.edu}
\affiliation{Department of Physics, Temple University, Philadelphia, Pennsylvania 19122, USA}
\author{Biswajit Santra}
\email{biswajit.santra@temple.edu}
\affiliation{Department of Physics, Temple University, Philadelphia, Pennsylvania 19122, USA}
\author{Puskar Bhattarai}
\affiliation{Department of Physics, Temple University, Philadelphia, Pennsylvania 19122, USA}
\author{Chandra Shahi}
\affiliation{Department of Physics, Central Michigan University, Mount Pleasant, MI, 48859, USA}
\author{Mark R. Pederson}
\affiliation{Department of Physics, University of Texas at El Paso, El Paso, TX, 79968, USA}
\author{Koblar A. Jackson}
\affiliation{Department of Physics, Central Michigan University, Mount Pleasant, MI, 48859, USA}
\author{John P. Perdew}
\affiliation{Department of Physics, Temple University, Philadelphia, Pennsylvania 19122, USA}
\affiliation{Department of Chemistry, Temple University, Philadelphia, Pennsylvania 19122, USA}

%\author{A. Author}
%\author{B. Author}%
% \email{Second.Author@institution.edu.}
%\affiliation{ 
%Authors' institution and/or address%\\This line break forced with \textbackslash\textbackslash
%}%

% \homepage{http://www.Second.institution.edu/~Charlie.Author.}
%\affiliation{%
%Second institution and/or address%\\This line break forced% with \\
%}%

\date{\today}% It is always \today, today,

\begin{abstract}
We study the importance of self-interaction errors in density functional approximations for various water--ion clusters. 
We have employed the Fermi-L\"owdin orbital self-interaction correction (FLOSIC) method in conjunction with LSDA, PBE, and SCAN to describe binding energies of hydrogen-bonded water--ion clusters, \textit{i.e.}, water--hydronium, water--hydroxide, water--halide, as well as non-hydrogen-bonded water--alkali clusters. 
In the hydrogen-bonded water--ion clusters, the building blocks are linked by hydrogen atoms, although the links are much stronger and longer-ranged than the normal hydrogen bonds between water molecules, because the monopole on the ion interacts with both permanent and induced dipoles on the water molecules.
We find that self-interaction errors overbind the hydrogen-bonded water--ion clusters and that FLOSIC reduces the error and brings the binding energies into closer agreement with higher-level calculations.
The non-hydrogen-bonded water--alkali clusters are not significantly affected by self-interaction errors.
Self-interaction corrected PBE predicts the lowest mean unsigned error in binding energies ($\leq$ 50 meV/\ce{H2O}) for hydrogen-bonded water--ion clusters.
Self-interaction errors are also largely dependent on the cluster size, and FLOSIC does not accurately capture the subtle variation in all clusters, indicating the need for further refinement.
\end{abstract}

\maketitle

%\begin{quotation}
%The ``lead paragraph'' is encapsulated with the \LaTeX\ 
%\verb+quotation+ environment and is formatted as a single paragraph before the first section heading. 
%(The \verb+quotation+ environment reverts to its usual meaning after the first sectioning command.) 
%Note that numbered references are allowed in the lead paragraph.
%%
%The lead paragraph will only be found in an article being prepared for the journal \textit{Chaos}.
%\end{quotation}

\section{Introduction}

Interactions between ions and water molecules play a critical role in many areas of physical chemistry, including electrochemistry, environmental chemistry, and biochemistry. Such water--ion interactions primarily involve hydrogen bonds (HBs), which appear with a wider range of bond strength (0.2--1.5 eV/\ce{H2O})~\cite{Meot-NerMautner2005} and can be much stronger than HBs within neutral water clusters~\cite{sharkas2020self}.
One of the most common ionic HBs occurs between water molecules and their self-ionized forms, \textit{i.e.}, H$_{3}$O$^{+}$ and OH$^{-}$, which are responsible for acid-base chemistry, proton transfer, etc. Despite a long history of research, the microscopic details of the mechanism behind the proton transfer dynamics is yet to be settled~\cite{Chen2017}.
Other ions, including halide ions and alkali ions, form strong bonds with water, and these interactions are also crucial for biochemistry and electrochemistry~\cite{Vegt2016}.

Understanding the molecular-scale structure and dynamics of aqueous ionic solutions requires concerted efforts from both experiments and computer simulations~\cite{Ohtaki1993,Marcus2009,Agmon2016,Vegt2016}.
Among the simulation methods, Kohn-Sham density functional theory (DFT)~\cite{DFT-Kohn-Sham} based \textit{ab initio} molecular dynamics is widely used and has been crucial in finding ion-solvation shell structures, autoionization processes, proton transfer mechanisms, and simultaneously the details of the electronic structure~\cite{Marx1999,geissler2001,tuckerman2002nature,chen2018hydroxide,DiStasio2014,Santra2015,Bankura2015, wang2011density}. 
%
% As much as the gas-phase water--ion clusters play a critical role in environmental chemistry, they are also important to benchmark the energetics, to explore the potential energy surface, and to understand the nature of water--ion interactions by energy decomposition~\cite{egan2020nature,egan2018assessing,egan2019assessing,Paesani2019,bizzarro2019nature}. 
\textcolor{black}{The gas-phase water--ion clusters are important to} benchmark the energetics, to explore the potential energy surface, and to understand the nature of water--ion interactions by energy decomposition~\cite{egan2020nature,egan2018assessing,egan2019assessing,Paesani2019,bizzarro2019nature}. 
In this work, we focus on the energetics of various gas-phase water--ion clusters.

Historically, the accuracy of describing HBs is not satisfactory with the most commonly used DFT exchange-correlation (XC) functionals (see review in ref.\citenum{Gillan2016} and references therein). 
The accuracy of DFT functionals for HBs is widely tested in neutral water clusters. One of the key failures of DFT is that common semi-local and hybrid XC functionals predict an incorrect energetic ordering of isomers of \textcolor{black}{the} water hexamer~\cite{Gillan2016,santra2008}. 
The origin of this error is mainly due to the lack of non-local van der Waals (vdW) interactions in those functionals~\cite{Gillan2016,santra2008}. 
The non-empirical strongly constrained and appropriately normed (SCAN)~\cite{PhysRevLett.115.036402} meta-GGA functional, which satisfies all 17 exact physical constraints that a semi-local XC functional can satisfy and includes intermediate-range dispersion, reproduces the correct energetic ordering among the isomers of the water hexamer~\cite{Sun2016,sharkas2020self}.
The strength of HBs also suffers from the self-interaction error (SIE) in DFT, and like all semi-local functionals SCAN is not free of SIE. 
Typically, the HBs in neutral water clusters are too strong due to the SIE~\cite{sharkas2020self,Gillan2016}.
The water--ion interactions are also affected by SIE, resulting in too strong binding energies of water--ion clusters with semi-local functionals~\cite{egan2020nature,egan2018assessing,egan2019assessing,Paesani2019,bizzarro2019nature}.
%

%\textcolor{red}{Identifying and fixing errors in specific quantum methods for applications to solutions containing anionic or cationic regions is challenging since the degree of electron delocalization impacts the monopole-dipole and monopole induced dipole energies which change significantly if the size or placement of a region containing excess charge is incorrect due to either errors of a quantum theory or quantum implementation. In other words, overly delocalized or misplaced excess charge changes the extent of weak bonds in all regions of a solution which in turn challenges a qualitative picture that attempts to account for hydrogen bonding in terms of relatively short-range multipolar induced dipole attractions between neighboring water molecules.}

Fixing errors in specific methods for systems containing anionic or cationic species is challenging since the degree of electron delocalization impacts the monopole-dipole and monopole induced dipole energies which change significantly if the size or placement of an excess charge containing species is incorrect due to either errors of the theory or the size of basis sets.
Among the systems studied here, the molecular anions OH$^{-}$ and OH$^{-}$(H$_{2}$O) and the atomic anions F$^{-}$, Cl$^{-}$, and Br$^{-}$ cannot bind the full excess electron in LSDA, PBE, and SCAN at the basis-set limit, while the other systems can. \textcolor{black}{The} SIE of semi-local density functional approximations tends to excessively delocalize electron density, which is more severe in small negatively-charged ions and molecules~\cite{PerZun-PRB-81}, typically leading to \textcolor{black}{non-negative values} for the Kohn-Sham eigenvalue of the highest occupied molecular orbital (HOMO)~\cite{Shore1977}.
The binding of the extra electron in anionic systems is also dependent on the choice of basis sets for electronic wavefunctions~\cite{Rosch1997,Rienstra2002,Jensen2010,Kim2011}.
A fully converged basis may \textcolor{black}{predict that a fraction of the extra electron is lost} to the continuum~\cite{Wasserman2017}, whereas, a localized moderate-sized basis can artificially confine the extra electron~\cite{Rosch1997,Rienstra2002,Jensen2010,Kim2011}\textcolor{black}{, or trap a fraction of it on another center \cite{ruzsinszky2006spurious}.}
Thus, self-interaction correction (SIC) to XC functionals is essential to achieve an accurate description of the density and energies of negatively-charged systems.

Hybrid functionals can partially alleviate SIE and predict more accurate HBs than semi-local functionals~\cite{Gillan2016, Santra2009}.
In comparison, the Perdew-Zunger (PZ) SIC approach offers a fully nonlocal orbital-by-orbital removal of electron self-interaction from local and semi-local XC functionals.
The PZ SIC method aims at eliminating the source of SIE in any approximate XC functional: its imperfect cancellation between the electron-electron self-repulsion (Hartree) and the approximate self-XC energy for any one-electron density~\cite{PerZun-PRB-81}.
Recently, the Fermi-L\"owdin orbital self-interaction correction (FLOSIC)~\cite{PedRuzPer-JCP-14,YanPedPer-PRA-17} was introduced as an efficient and unitarily invariant approach for implementing PZ SIC that can be used in conjunction with any approximate XC functional.
The computational effort in the FLOSIC method allows for its application to clusters that were inaccessible through previous implementations~\cite{Pederson1985,Vydrov2004} of PZ SIC.
The FLOSIC methodology has been used to study properties of chemical and physical interest for a range of systems~\cite{Pederson2016,Hahn2017,Kao2017,Sharkas2018,Joshi2018,Shahi2019,Johnson2019,Withanage2019,Li2020,Adhikari2020} including water~\cite{sharkas2020self,Vargas2020,Akter2020,Batool2019}.
It has been shown that PZ SIC applied with SCAN significantly reduces the overbinding of HBs in neutral water clusters, predicting binding energies in much closer agreement with the CCSD(T)-F12, and simultaneously retaining the correct energetic ordering among the water hexamer isomers~\cite{sharkas2020self}.
It is thus quite important to explore how PZ SIC in conjunction with SCAN or other functionals performs in water--ion interactions.

Here, we have studied the effect of removing SIE on three non-empirical XC functional approximations, namely the local spin-density approximation (LSDA), the generalized gradient approximation formulated by Perdew, Burke, Ernzerhof (PBE)~\cite{perdew1996}, and the SCAN meta-GGA functional \cite{PhysRevLett.115.036402}, in predicting the binding energies of protonated water clusters [H$_{3}$O$^{+}$(H$_2$O)$_{n}$, $n$=1,2,3 and 6], deprotonated water clusters [OH$^{-}$ (H$_{2}$O)$_{n}$, $n$=1--6],
water--halide ion clusters [X$^{-}$(H$_{2}$O)$_{n}$, X=(F, Cl, Br), $n$=1--2], and water--alkali ion clusters [M$^{+}$(H$_{2}$O)$_{n}$, M=(Li, Na, K), $n$=1--2]. 
We have analyzed how first-order and second-order (density--driven) errors in  water--ion clusters depend on SIE.
We find that self-interaction errors overbind the hydrogen-bonded water--ion clusters, and that FLOSIC reduces the error and brings the binding energies into closer agreement with higher-level calculations.
FLOSIC applied to PBE predicts the lowest mean error in binding energies ($\leq$ 50 meV/\ce{H2O}) for hydrogen-bonded water--ion clusters.

\begin{figure*}[htb]
%\centering  
\includegraphics[width=0.95\linewidth]{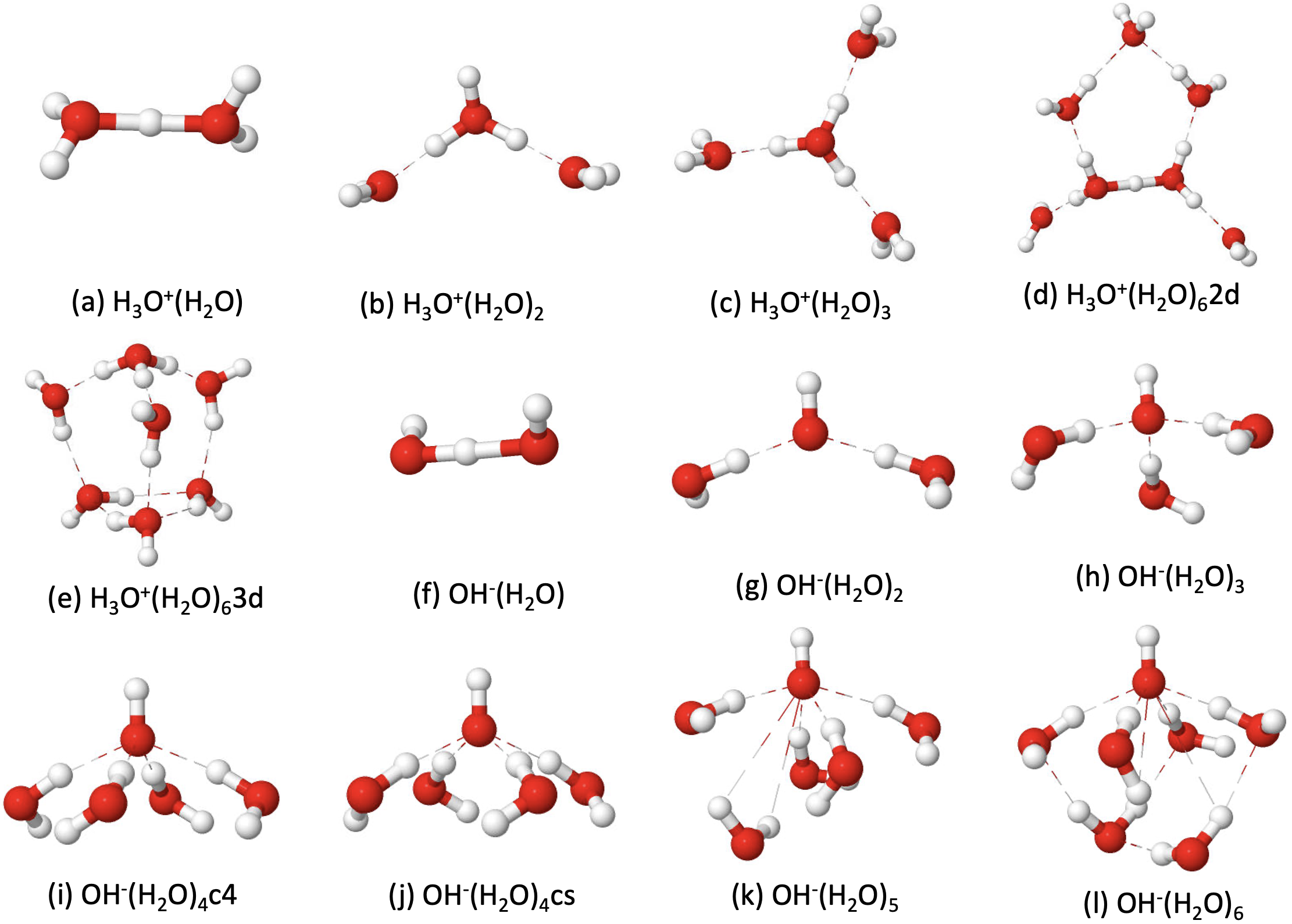}
\caption{Structures of protonated and deprotonated water clusters. Five protonated water clusters H$_{3}$O$^{+}(\text{H}_2\text{O})_n$ ($n$ = 1--3, and 6) and seven deprotonated water clusters OH$^{-} (\text{H}_2\text{O})_n$ ($n$ = 1--6) are taken from Ref.\citenum{doi:10.1021/ct800549f}. The red and white spheres respectively indicate oxygen and hydrogen atoms. The dashed lines represent hydrogen-bonds.}
\label{H2OCLUSTERS}
\end{figure*}

\begin{figure*}[htb]
\includegraphics[width=0.80\linewidth]{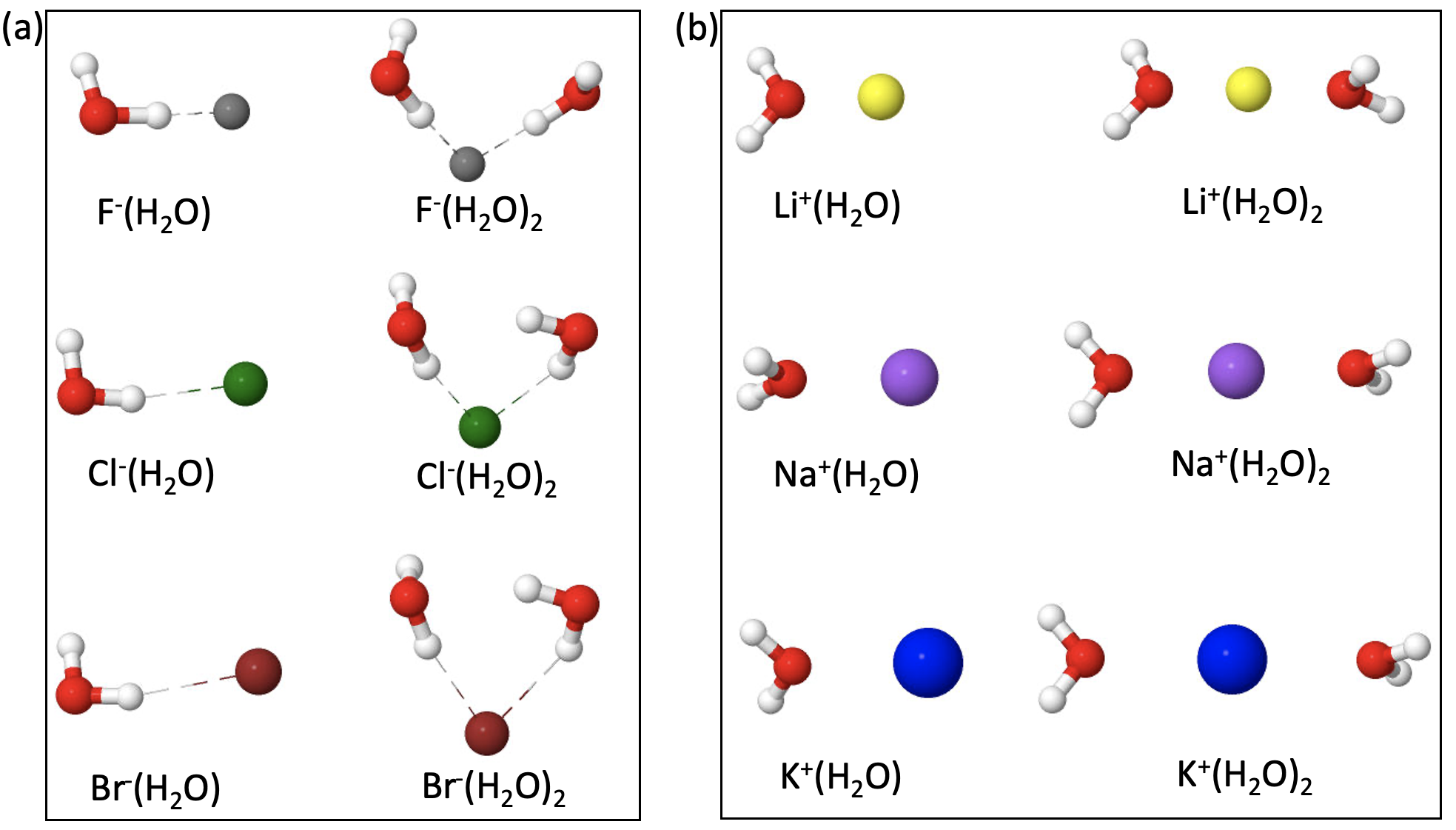}
\caption{Structures of the studied (a) water--halide clusters and (b) water--alkali clusters. Oxygen and hydrogen are represented by red and white spheres. The anions and cations are marked by other colors. Dashed lines represent hydrogen-bonding.}
\label{halide-str}
\end{figure*}

%%%%%%%%%%%%%%%%%%%%%%%%%%%%%%%%
\section{Methodology}
%%%%%%%%%%%%%%%%%%%%%%%%%%%%%%%
\subsection{DFT and FLOSIC Calculations}

Here, all calculations with density functional approximations (DFA) and the corresponding FLOSIC applied functionals, FLOSIC--DFA, were performed with the all-electron FLOSIC code~\cite{pederson1990variational,FLOSIC-GIT}, which was developed from the parent NRLMOL density--functional code~\cite{JacPed-PRB-90,PedJac-PRB-91,PorPed-PRB-96,PRB.58.1786,PedPorKorPat-PSSB-00}.
The code uses contracted Gaussian--type orbitals as basis sets that are optimized for density functionals and referred to as the DFO~\cite{PorPed-PRA-99}.
% and integrates Hamiltonian matrix elements numerically on a variational mesh\cite{PedJac-PRB-90}%. 
We have performed systematic basis-set convergence tests by adding diffuse functions on top of the DFO basis set. The details regarding the construction of the diffuse functions and basis set convergence are discussed in the next section. 
%Spin unpolarized, non--relativistic all--electron KS--DFT calculations were carried out using 
We used the local--density approximation (LSDA) which consists of the Slater--Dirac exchange\cite{Sla-PR-51} and Perdew--Wang 92\cite{PerWanC-PRB-92} correlation functional, the semi-local PBE\cite{PerBurErn96} and the SCAN\cite{PhysRevLett.115.036402} XC functionals.
All calculations were done using an accurate integration grid \cite{pederson1990variational}. For calculations involving the SCAN functional, especially dense grids were used~\cite{yamamoto2019fermi}.

The FLOSIC methodology\cite{PedRuzPer-JCP-14,YanPedPer-PRA-17} is based upon the original PZ--SIC\cite{PerZun-PRB-81} method. In this method, SIE is removed from an approximate functionals 
%$E_{\text{xc}}^{\text{approx}}[n_{\uparrow}(\b{r}), n_{\downarrow}(\b{r})]$ 
$E_{\text{xc}}^{\text{approx}}[n_{\uparrow}, n_{\downarrow}]$ 
orbital by orbital, as given by
\begin{align}
\label{EPZ}
    E^{\text{PZ--SIC}}& = E_{\text{xc}}^{\text{approx}}[n_{\uparrow},n_{\downarrow}] \, - \nonumber \\ 
    &\sum_{i\sigma}(E^{\text{approx}}_{\text{xc}}[n_{i \sigma},0]
    + U[n_{i \sigma}]),
\end{align}
where $n_{i \sigma}$ is a single orbital density and $U[n_{i \sigma}]$ is the Hartree electrostatic energy of that one-electron density.
The orbitals must be localized to make $E^{\text{PZ--SIC}}$ size-extensive~\cite{PerZun-PRB-81,pederson1988localized}.
%Perdew and Zunger further made the important conclusion that, SIC orbitals should be localized by a unitary transformation of the Kohn-Sham orbitals to achieve size consistency. Later, Pederson and collaborators performed the SIC calculations for the first time based upon such a unitary transformation \cite{pederson1988localized}. 
The search for sufficiently localized orbitals that minimize Eq.~\ref{EPZ} becomes computationally efficient with Fermi-Orbitals (FOs)\cite{pederson1988localized}. The FOs~\cite{LukBer82,LukCul84} are obtained as
\begin{equation}
F_{i \sigma}({\textbf{r}}) = \frac{\sum_{j}^{N_{\sigma}} \psi_{j \sigma}^{*}({\textbf{a}_{i \sigma})}  
\psi_{j \sigma}(\textbf{r})}{\sqrt{\sum_{j}^{N_{\sigma}}|\psi_{j \sigma}(\textbf{a}_{i \sigma})|^{2}}},
\end{equation}
where $N_{\sigma}$ is the number of electrons with spin $\sigma$. Each FO, $F_{i \sigma}({\textbf{r}})$, depends on a position vector, $\textbf{a}_{i \sigma}$, called a Fermi--Orbital Descriptor (FOD), and the spin density constructed by any set of orthonormal orbitals, $\psi_{j \sigma}(\textbf{r})$, that spans the occupied space. The localized FOs are normalized but not mutually orthogonal. The L\"{o}wdin method of symmetric orthogonalization \cite{Low-JCP-50} is then performed to obtain the orthonormal Fermi-L\"{o}wdin orbitals (FLOs).

We employed a fully self-consistent procedure~\cite{YanPedPer-PRA-17} to minimize the FLOSIC--DFA total energy. The calculations were initialized with a set of guessed FODs~\cite{schwalbe2019interpretation} to obtain self-consistent FLOs and total energy using an energy tolerance of $10^{-7}$ Hartree. Then the FODs are updated using derivatives of total energy with respect to FOD positions~\cite{pederson2015,PedersonBaruah2015,hahn2015fermi} and gradient optimization methods, starting from the scaled~\cite{Jackson_2019} limited memory Broyden--Fletcher--Goldfarb--Shanno (L--BFGS)\cite{BFGS1,BFGS2,BFGS3,BFGS4} and switching, if necessary, to the conjugate gradient. Then self-consistent FLOs are obtained with the new set of FODs. The iterative minimization of the total energy with respect to FODs is performed until the maximum FOD force component drops below $10^{-3}$ Hartree/Bohr.

%FLOSIC calculations performed in conjunction with the uncorrected density functional approximations (DFAs) are referred to as FLOSIC--DFA. 
We have also computed DFA energies on the corresponding FLOSIC--DFA self-interaction corrected density and FLOs, which is referred to as DFA@FLOSIC.
%LSDA, PBE, and SCAN one-shot (non-self--consistent) energies evaluated on the self--consistent FLOSIC--LSDA, FLOSIC--PBE, and FLOSIC--SCAN densities, respectively will be named LSDA(@FLOSIC), PBE(@FLOSIC), and SCAN(@FLOSIC). 
The DFA@FLOSIC energies enable us to quantify the magnitude of the error in energy coming from a DFA density that suffers from self-interaction. The difference between DFA and DFA@FLOSIC energies is analogous to the commonly known density--driven--error~\cite{Kim2013,kim2015improved}. The Hartree-Fock density is often used to quantify the density--driven error since it is self-interaction (exchange only) free~\cite{Kim2013,kim2015improved}. The FLOSIC--DFA densities are self-interaction (exchange-correlation) free by definition.

\subsection{Reference Geometry and Binding Energy}

Fig.~\ref{H2OCLUSTERS} shows structures of protonated and deprotonated clusters, which include five protonated water clusters H$_{3}$O$^{+}(\text{H}_2\text{O})_n$, $n$=1--3, and 6, and seven deprotonated water clusters OH$^-$(H$_2$O)$_n$, $n$=1--6.
The nuclear coordinates of those clusters are taken from the WATER27\cite{doi:10.1021/ct800549f} set which is part of the general main group thermochemistry, kinetics, and non-covalent interactions (GMTKN55)\cite{GoeHanBauEhrNajGri-PCCP-17} benchmark database. 
The WATER27 geometries were optimized \cite{GoeHanBauEhrNajGri-PCCP-17} at the B3LYP\cite{PhysRevA.38.3098,PhysRevB.37.785}/6-311++G(2d,2p) level of theory. 
Fig.~\ref{halide-str} shows the structures of water--halide and water--alkali clusters.
We have considered six water--halide clusters X$^{-}$(H$_{2}$O)$_{n}$ with X=(F, Cl, Br) and ($n$=1 and 2) and six water--alkali clusters M$^+$(H$_{2}$O)$_n$ with M=(Li, Na, K) and ($n$=1 and 2)
The nuclear coordinates of those clusters are taken from Ref.\citenum{bizzarro2019nature,egan2020nature} which were optimized at the DF-MP2 \cite{werner2003fast} level of theory with an aug-cc-pVTZ basis set.

For the clusters, the binding energy per water molecule ($E_b$) is defined as
\begin{equation}
 E_{b} = \frac{E^{\text{cluster}} - n E^{\rm H_2O}-E^{\rm ion}}{n} \,,
\label{Eb}
\end{equation}
where $E^{\text{cluster}}$ is the total energy of a cluster that includes $n$ H$_2$O molecules and one ion. $E^{\rm H_2O}$ is the total energy of an isolated H$_2$O monomer at its equilibrium geometry and $E^{\rm ion}$ is the total energy of an isolated ion. $E_b$ is negative for a stable cluster.
Reference binding energies of protonated and deprotonated water clusters were computed at  the explicitly correlated coupled-cluster level of theory, CCSD(T)-F12b~\cite{Adler2007,knizia2009simplified}, at the complete basis set (CBS) limit, and are taken from Ref.\citenum{Manna2017}.
The reference binding energies with CCSD(T)-F12b at the CBS limit are taken from Ref.\citenum{egan2020nature} for the water--halide and from Ref. \citenum{arismendi2016ttm} for the water--alkali clusters.
%
%For halide-water clusters, the reference CCSD(T) binding energies are obtained from the authors from the same work~\cite{egan2020nature} upon request.

%The ion is hydronium ion (H$_{3}$O$^{+}$) in protonated water clusters, hydroxyl ion (OH$^{-}$) in deprotonated water clusters, halide ion (X$^{-}$) in halide-water clusters, and alkali ion (M$^{+}$) in alkali ion-water clusters.\\

\begin{figure*}[htb]
\includegraphics[width=1.0\linewidth]{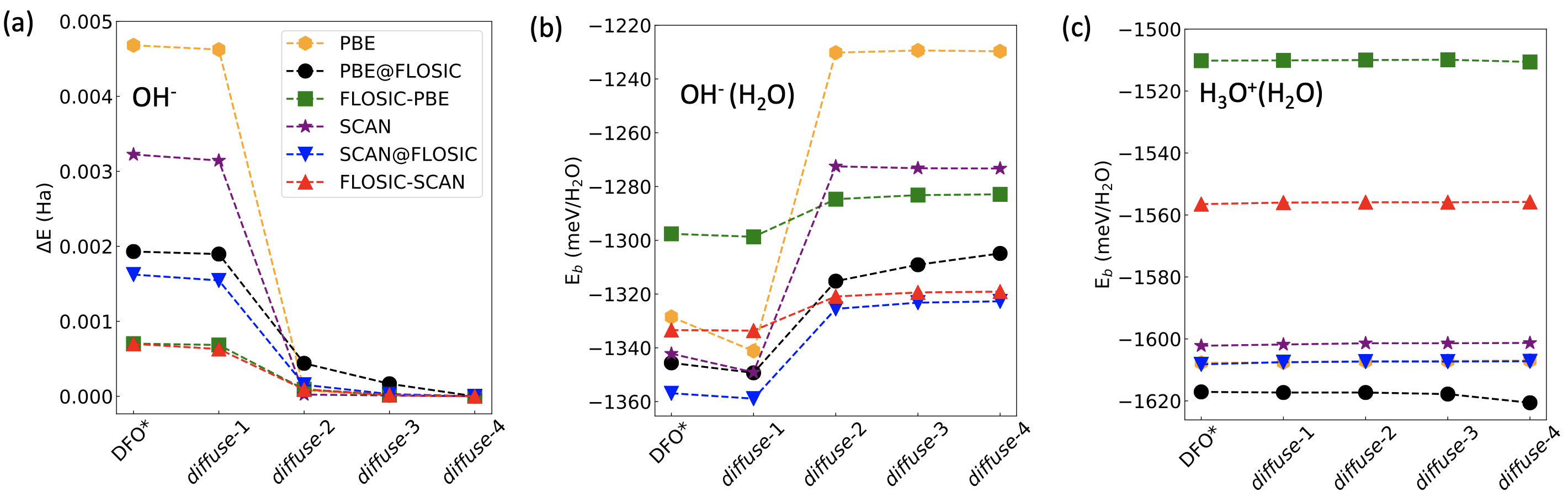}
\caption{(a) The basis-set dependence of the total energy of OH$^{-}$, plotted with respect to the \textit{diffuse}-4 basis (see text for the details of basis). Variation of the binding energy (E$_b$) of the (b) \ce{OH^-(H2O)} and (c) \ce{H3O^+(H2O)} molecules with increasing size of the basis. Here DFA@FLOSIC refers to the DFA energy computed at the corresponding FLOSIC--DFA density. At the complete basis set limit, only SIC can bind the full excess electron in OH$^{-}$ and \ce{OH^-(H2O)}.}
\label{DDE}
\end{figure*}

\section{Results and Discussions}
\subsection{Basis-Set and Density Error}

Given that the semi-local functionals LSDA, PBE, and SCAN can bind only a fraction of the excess electron in the complete-basis-set limit as discussed in the introduction, we will use for those functionals basis sets without diffuse basis functions. Restricting the basis set is a familiar computational choice that makes the semi-local functionals look somewhat better than they are. The rest of this section will discuss the basis sets and their effects without and with self-interaction.

The erroneous delocalization of the extra electron in anions obtained with semi-local DFAs is dependent on the size of the basis set.
On the one hand, the use of a moderate-sized localized basis can compensate for the delocalization. 
On the other hand, accurate description of HBs often requires a large basis with some diffuse Gaussian functions (\textit{i.e.},  with relatively small exponents).
The deprotonated water clusters are negatively charged and held together by HBs, for which the need for diffuse functions is ambiguous, and we explore this issue here. 
Fortunately, through the FLOSIC--DFA method we have electron densities of the negatively charged clusters that are free from self-interaction and that can be used to test the energy convergence with the size of the basis.

First, we briefly describe the basis sets we have used to determine the energy convergence.
The NRLMOL DFO basis with standard additional polarization functions consists of 5$s$,4$p$,4$d$ functions for O and 4$s$,4$p$,2$d$ for H, which we refer to as DFO$^\ast$.
On top of the DFO$^\ast$ basis, we add diffuse functions of types $s$, $p$, and $d$ for both O and H. 
The modified basis sets are defined as: i) \textit{diffuse}-1 which includes O($s$) and H($s$) diffuse functions, ii) \textit{diffuse}-2 which includes O($s$,$p$) and H($s$) diffuse functions, iii) \textit{diffuse}-3 which includes O($s$,$p$) and H($s$,$p$) diffuse functions, and iv) \textit{diffuse}-4 which includes O($s$,$p$,$d$) and H($s$,$p$) diffuse functions.

Fig.~\ref{DDE}(a) shows the convergence of the total energy of \ce{OH^-} with respect to an increasing number of diffuse basis functions.
The energies are almost converged around the \textit{diffuse}-4 basis, and with respect to that, the energies obtained with DFO$^\ast$ differ by less than 1 mHa in both FLOSIC--PBE and FLOSIC--SCAN.  
In contrast, the two semi-local functionals show a much larger change in energy than their FLOSIC counterparts, and a significant portion of this discrepancy is due to erroneous density since PBE and SCAN energies computed at their corresponding self-interaction corrected densities, \textit{i.e.}, PBE@FLOSIC and SCAN@FLOSIC, substantially reduce the discrepancy. \textcolor{black}{For this small anion, in PBE or SCAN 
without SIC, a fraction of the excess electron would escape in the complete basis-set limit.}
%
%The FLOSIC--DFA self-consistent electron densities eliminate it to a large extent and are taken to be "exact" in this study.
%
Here we regard the difference between DFA energy computed at its self-consistent density and self-interaction free FLOSIC--DFA density as \textcolor{black}{an estimate of the} density-driven error.
The density-driven error is quite significant in \ce{OH^-} (where it could be even larger than we have estimated here) and is negligible in \ce{H3O^+} and \ce{H2O} molecules.
This indicates that extra diffuse functions can exaggerate the erroneous density delocalization in \ce{OH^-} and a suitable choice of a converged basis is not straightforward for PBE and SCAN.

\begin{figure}[htb]
\includegraphics[width=0.95\linewidth]{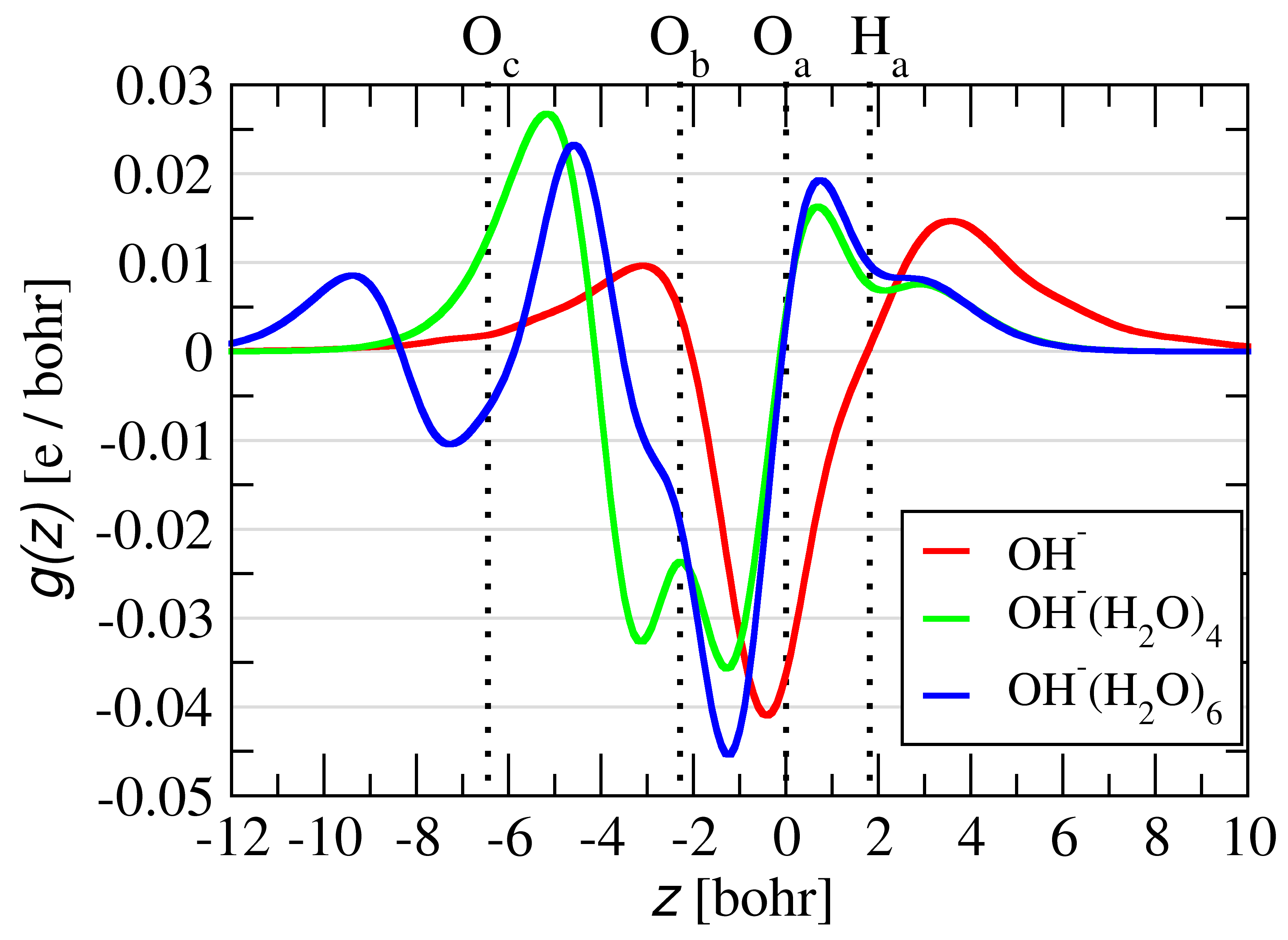}
\caption{Projection onto the OH$^-$ axis of the PBE self-interaction error of the electron density for the hydroxyl group and two hydroxyl-water clusters. For further explanation, see the last two paragraphs of section 3.1.
}
\label{fig-4}
\end{figure}

\begin{table}[htb]
 \caption{Three measures of the size of the PBE self-interaction error of the electron density in hydroxyl-water clusters. For further explanation, see the last two paragraphs of section 3.1.}
\begin{ruledtabular}
  \begin{tabular}{cccc}
    System & $\delta n_{z>0}$ (e)& $\delta \mu_z$ (D) & $|\delta \mu|$ (D) \\
   \hline
   \ce{OH^-}          & 0.026 &  0.37 & 0.37 \\
   \ce{OH^-(H2O)4}    & 0.042 & -0.07 & 0.09 \\
   \ce{OH^-(H2O)6}    & 0.048 &  0.02 & 0.11 \\
  \end{tabular}
\end{ruledtabular}   
\label{tb2}
\end{table}

Fig.~\ref{DDE}(b) shows the convergence of the binding energy of \ce{OH^-(H2O)} with respect to basis. 
From DFO$^\ast$ to \textit{diffuse}-3 basis $E_b$ changes by $\sim$14 meV/\ce{H2O} with both FLOSIC--PBE and FLOSIC--SCAN, but the change is much larger with PBE (99 meV/\ce{H2O}) and SCAN (69 meV/\ce{H2O}).
This again shows that the addition of diffuse functions for negatively charged molecules can be problematic without SIC.
%
%The self-interaction corrected density allows to determine binding energies with self-interaction uncorrected functionals with better accuracy.
%
The PBE@FLOSIC and SCAN@FLOSIC show a much smoother convergence of $E_b$, predicting, respectively, a 40 meV/\ce{H2O} and 33 meV/\ce{H2O} difference between the DFO$^\ast$ and \textit{diffuse}-3 basis.
The density-driven error (\textit{i.e.}, the difference between DFA and DFA@FLOSIC) in the binding of \ce{OH^-(H2O)} becomes much larger with additional diffuse functions, more so in PBE (76 meV/\ce{H2O}) than in SCAN (51 meV/\ce{H2O}) with the \textit{diffuse}-3 basis.
In comparison, the $E_b$ of \ce{H3O^+(H2O)} shows very little ($<$10 meV/\ce{H2O}) density-driven error with and without the diffuse functions, as shown in Fig.~\ref{DDE}(c).
From Fig.~\ref{DDE}(b) we conclude that PBE@FLOSIC and SCAN@FLOSIC binding energies are converged at the \textit{diffuse}-3 basis, which is similar to the convergence of the FLOSIC--DFAs.
Also, Fig.~\ref{DDE}(b) shows that the PBE (SCAN) energy at the DFO$^\ast$ basis are close to the PBE@FLOSIC (SCAN@FLOSIC) energy at the \textit{diffuse}-3 basis, indicating that the error due to a relatively smaller basis cancels to some extent with the density delocalization errors.
In the following, we report and discuss results obtained with the DFO$^\ast$ for LSDA, PBE, and SCAN, and with the \textit{diffuse}-3 basis for all other functionals.

There will be a self-interaction (delocalization) error of the electron density from a density functional (\textit{e.g.}, PBE) approximation, even in anion clusters like \ce{OH^-(H2O)_$n\geq 2$} that bind the full extra electron in the complete basis-set limit for that approximation. Here we will show that this error is rather small but systematic, at least up to $n$=6.  Fig.~\ref{H2OCLUSTERS} shows that the hydroxyl anion is found on the surface of each of these clusters.  
It could be interesting in future work to see if the delocalization error is larger in larger clusters or bulk, where the hydroxyl anion can be completely solvated by the waters.

For this purpose, we define the PBE self-interaction error of the density as $\delta n{\rm (\textbf {r})}=n_{\rm PBE}{\rm (\textbf {r})} - n_{\rm FLOSIC-PBE}{\rm (\textbf {r})}$, a function that integrates to zero electrons. To visualize this, we take the origin of coordinates on the O$^-$ or O$_a$ nucleus, and the $z$ axis pointing to the H or H$_a$ nucleus of the hydroxyl group. Then we project $\delta n{\rm (\textbf {r})}$ onto the $z$ axis via
\begin{equation}
 g(z) = \int dx dy dz^\prime \delta n(x,y,z^\prime) \frac{a}{\sqrt \pi} \exp{-a^2(z^\prime - z)^2} ,
\label{eq-4}
\end{equation}
where $a$=1 bohr$^{-1}$ is large enough for atomic resolution and small enough for numerical real-space grids.  
The interpretation is that $\int_A^B dz g(z)$ is the PBE self-interaction error in the number of electrons between the planes $z=A$ and $z=B$.

\begin{table*}[htb]
 \caption{Binding energies with density functionals and the corresponding mean unsigned errors (MUEs) computed with respect to the CCSD(T)-F12b reference~\cite{Manna2017} for deprotonated and protonated water clusters. The final column shows the percent error (PE) for FLOSIC--PBE. The energies are obtained with the basis DFO$^\ast$ for LSDA, PBE, and SCAN, and with \textit{diffuse}-3 for all other functionals. The corresponding reference value for the neutral \ce{(H2O)6} prism is -332 meV/\ce{H2O}, suggesting that the charged group is bound strongly to all the water molecules, as sketched in Fig. 1.}
\begin{ruledtabular}
\begin{tabular}{lccccccccccc}  
    &     \multicolumn{10}{c}{Binding Energy (meV/H$_2$O)}& PE \\
    \cline{2-11}
%    Cluster & Ref.& \MyHead{1.2cm}{LSDA}
%      & \MyHead{1.2cm}{\relsize{-3} LSDA\\ {\relsize{-3}@FLOSIC}}
%      & \MyHead{1.2cm}{{\relsize{-3}FLOSIC}\\ \relsize{-3} --LSDA} & \MyHead{1.2cm}{PBE}& \MyHead{1.2cm}{\relsize{-3} PBE\\{\relsize{-3}@ FLOSIC}}& \MyHead{1.2cm}{{\relsize{-3}FLOSIC}\\\relsize{-3} --PBE}& \MyHead{1.2cm}{SCAN}& \MyHead{1.2cm}{\relsize{-3} SCAN\\{\relsize{-3}@FLOSIC}}& \MyHead{1.2cm}{{\relsize{-3}FLOSIC}\\\relsize{-3} --SCAN}&\MyHead{1.2cm}{{\relsize{-3}FLOSIC}\\\relsize{-3} --PBE} \\
    Cluster & Ref. & LSDA & LSDA@  & FLOSIC-- & PBE & PBE@   & FLOSIC-- & SCAN & SCAN@  & FLOSIC-- & FLOSIC-- \\
            &      &      & FLOSIC & LSDA     &     & FLOSIC & PBE      &      & FLOSIC & SCAN     & PBE      \\
%      
%    \cmidrule(rl){1-1}\cmidrule(rl){2-2}\cmidrule(rl){3-3}\cmidrule(rl){4-4}\cmidrule(rl){5-5}\cmidrule(rl){6-6}\cmidrule(rl){7-7}\cmidrule(rl){8-8}\cmidrule(rl){9-9}\cmidrule(rl){10-10}\cmidrule(rl){11-11}\cmidrule(l){12-12}
%    
\hline
     \ce{OH^{-}(H_{2}O)} & -1157&-1646&	-1631&	-1504&	-1329&	-1305&	-1283	&-1342&	-1324&	-1319&10.9 \%
 \\
     \ce{OH^{-}(H_{2}O)_{2}} &-1056&-1408	&-1399& -1299&	-1147&	-1136&	-1108	&-1176&	-1168&	-1147&5.0 \%
 \\
     \ce{OH^{-}(H_{2}O)_{3}} &-976& -1273&	-1272& -1213	&-1038&	-1038	&-1041&	-1069&	-1066	&-1066&6.7 \%
 \\
     \ce{OH^{-}(H_{2}O)_{4} c 4} &-915&-1175&	-1170&-1106&	-938&	-937&	-929&	-992&	-989&	-967&1.6 \%
 \\
     \ce{OH^{-}(H_{2}O)_{4} cs} &-922&-1235& -1229	&-1147&	-956&	-954&	-943	&-1013&	-1009&	-982&2.4 \%
 \\
     \ce{OH^{-}(H_{2}O)_{5}} &-874& -1168&	-1160&	-1078	&-890	&-888	&-873&	-955&	-953&	-921&-0.1 \%
 \\
     \ce{OH^{-}(H_{2}O)_{6}} &-836& -1134&	-1126&	-1037&-854	&-853&	-833&	-914&	-912&	-881&-0.4 \%
 \\
    \hline
    MUE   &--&329 &322 &235& 59& 54& 41 & 104 & 98 &	78&--
 \\
  \hline
  \ce{H_{3}O^{+}(H_{2}O)}&-1463& -1909	&-1926&	-1742&	-1608&	-1618	&-1510&	-1602	&-1608&	-1556&3.2 \% \\
    \ce{H_{3}O^{+}(H_{2}O)_{2}}&-1238&  -1569	&-1577&-1451	&	-1314&	-1323&	-1253&	-1330&	-1335&	-1276&1.2 \%
 \\
    \ce{H_{3}O^{+}(H_{2}O)_{3}}&-1110 & -1380&	-1387&	-1306&	-1156&	-1165&	-1131&	-1173&	-1179&	-1152&1.9 \%
 \\
    \ce{H_{3}O^{+}(H_{2}O)_{6} 2d}&-830& -1083&-1086&	-1008&	-865&	-872&	-843&	-882&	-885&	-855&1.6 \%
 \\
    \ce{H_{3}O^{+}(H_{2}O)_{6} 3d}&-851 & -1127	&-1129&	-1051&	-878&	-887&	-868&	-908&	-911&	-881&2.0 \%
 \\
    \hline
    MUE      &-- &315 &323 &213&	66	&75	&23 &81& 85&	46&--
 \\
  \end{tabular}
  \end{ruledtabular}
%  }
  \label{tab1}
\end{table*}

Fig.~\ref{fig-4} shows $g(z)$ for the hydroxyl anion and its bound complexes with four or six water molecules. The oxygens O$_b$ on the four nearest-neighbor waters are located at approximately the same negative $z$, and those O$_c$ on the two second-neighbor waters are located at a more negative $z$. (Average $z$ coordinates for each set of neighboring oxygens are shown in Fig.~\ref{fig-4}) The PBE self-interaction error clearly removes a small fraction of an electron from each oxygen site, and transfers it to the hydrogen sites or to the surface (and in particular the ends) of the cluster.
Table~\ref{tb2} shows the small fraction of an electron transferred into the half-space $z>0$ ($\delta n_{z>0}$), and the corresponding small PBE self-interaction errors in the $z$ component ($\delta \mu_z$) and magnitude of the dipole moment ($|\delta \mu|$). 
Note that the magnitudes of the dipole moment of OH$^-$ in our calculation using PBE and FLOSIC--PBE are 0.58 Debye and 0.95 Debye, respectively.
For comparison, the magnitudes of the dipole moment of an isolated water molecule are 1.79 Debye and 1.91 Debye, respectively from PBE and FLOSIC--PBE.

\begin{figure*}[htb]
\includegraphics[width=0.95\linewidth]{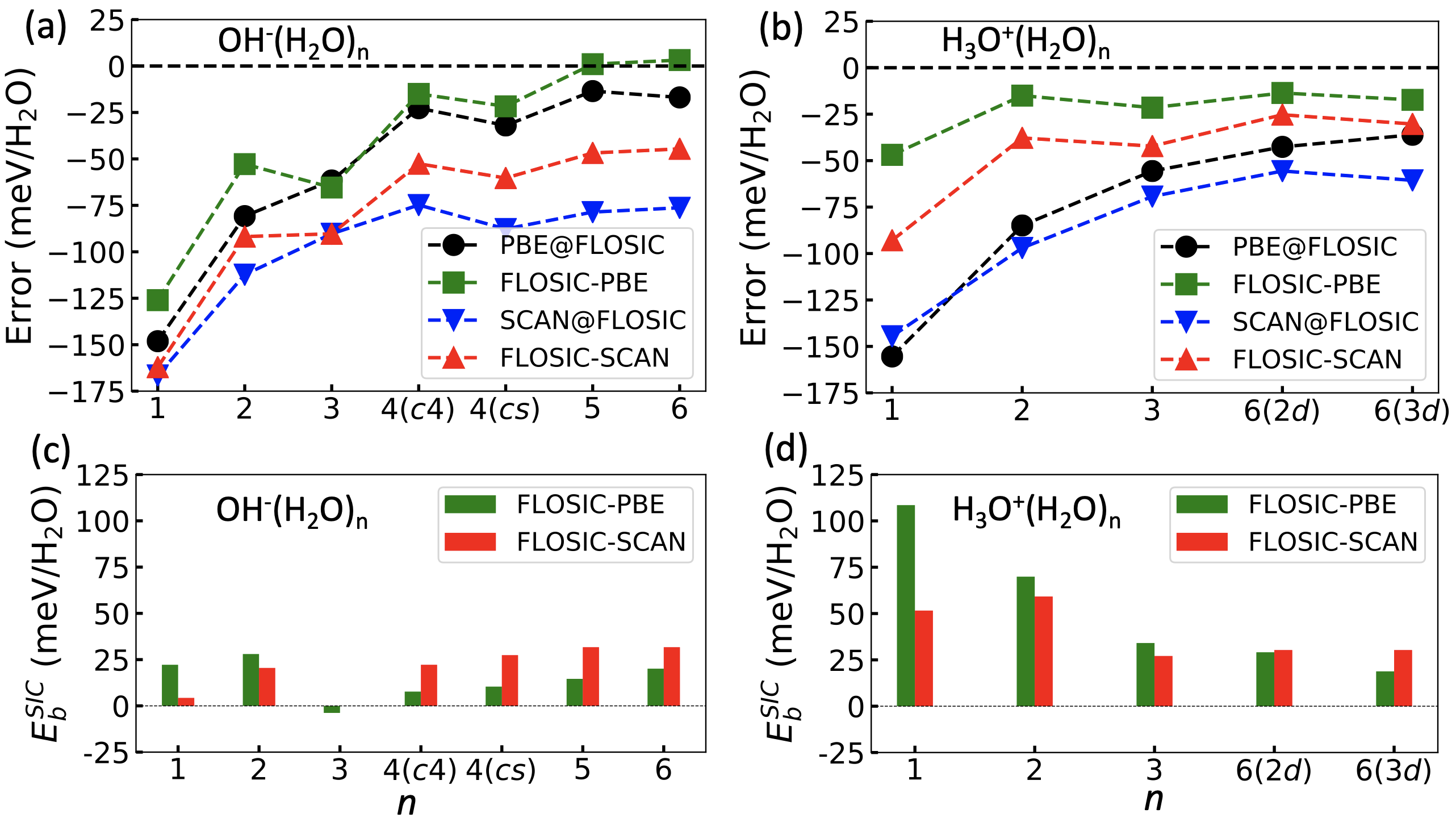}
\caption{Error in the binding energies of the (a) deprotonated water clusters OH$^{-} (\text{H}_2\text{O})_n$ ($n$ = 1--6) and (b) protonated water clusters H$_{3}$O$^{+}(\text{H}_2\text{O})_n$ ($n$ = 1--3, and 6) with various functionals compared to the CCSD(T)-F12b reference~\cite{Manna2017}. Here a negative sign indicates overbinding. Self-interaction correction to the binding energies ($E_b^{\rm SIC}$) of the (c) deprotonated and (d) protonated water clusters.}
\label{wrt-size}
\end{figure*}

\subsection{Deprotonated and Protonated Water Clusters}

The hydroxide and hydronium ions form strong hydrogen bonds with water molecules.
It can be seen from the reference~\cite{Manna2017} CCSD(T)/F12b energies that the binding energy of \ce{H3O^+(H2O)} is stronger by 306 meV/\ce{H2O} stronger than that of \ce{OH^-(H2O)} (Table~\ref{tab1}).
In \ce{H3O^+(H2O)}, the shared proton sits exactly at the center between the two oxygen atoms separated by 2.40 \AA\ (Fig.~\ref{H2OCLUSTERS}(a)), whereas, in \ce{OH^-(H2O)}, the O-O distance is 2.47 \AA, and the shared proton has one covalent O-H bond of length 1.12 \AA\ and one hydrogen bond of length 1.34 \AA.
In the bigger clusters, the binding per molecule decreases due to relatively weak water--water interactions, $\sim$-330 meV/H$_{2}$O for (H$_{2}$O)$_{6}$~\cite{sharkas2020self}.
With increasing cluster size, the binding per molecule gradually decreases, as explained at the end of the caption of Table 1.
It is interesting that both H$_3$O$^+$(H$_2$O)$_6$ and OH$^-$(H$_2$O)$_6$ clusters become almost equally stable, differing by $<$15 meV/\ce{H2O} with CCSD(T)/F12b.
All of the studied XC functionals qualitatively reproduce the relative stability of these two types of clusters (Table~\ref{tab1}).  
However, there are non-negligible errors in the binding energy of an individual cluster obtained with all functionals. 
We explore the errors in more detail in Fig.~\ref{wrt-size}, omitting the LSDA-based methods which provide errors $>$200 meV/\ce{H2O}. 

Overall, it is evident from Fig.~\ref{wrt-size}(a) and (b) that FLOSIC--PBE performs better than FLOSIC--SCAN for almost all of these clusters.
The MUEs indicate that there is a systematic reduction in the error in binding energies with the FLOSIC--DFA methods compared to the corresponding parent DFAs (Table~\ref{tab1}).
%
%LSDA@FLOSIC predicts on average too strong ($\sim$322 meV/\ce{H2O}) binding energies for both types of clusters. 
%
%FLOSIC--LSDA largely reduces these errors but still concedes sizable MUEs of $>$200 meV/\ce{H2O}.
%
%At the level of semi-local DFAs improvements in the MUEs are already quite significant compared to LSDA. 
%
%Compared to LSDA both PBE@FLOSIC and SCAN@FLOSIC predict much smaller MUEs of $<$100 meV/\ce{H2O}.
%
%On average self-interaction correction weakens the binding energies and brings in a closer agreement with CCSDT(T)/F12b.
%
For the deprotonated clusters, the MUE of FLOSIC--SCAN is 78 meV/\ce{H2O}, reducing the MUE by 20 meV/\ce{H2O} from SCAN@FLOSIC.
For the protonated clusters, SCAN@FLOSIC and FLOSIC--SCAN predict MUEs of 85 meV/\ce{H2O} and 46 meV/\ce{H2O}, respectively.
FLOSIC--PBE provides the best agreement with the reference binding energies in both types of clusters, predicting MUEs of 41 meV/\ce{H2O} and 23 meV/\ce{H2O} for deprotonated and protonated clusters, respectively.
We find that our FLOSIC--PBE MUE for the deprotonated clusters is comparable to the hybrid functional PBE0/def2-QZVP mean error (31 meV/\ce{H2O}) reported by Grimme and co-workers~\cite{goerigk2017look}.
In addition, the FLOSIC--PBE MUE for the protonated clusters is more accurate than the PBE0/def2-QZVP method (MUE of 44 meV/\ce{H2O})~\cite{goerigk2017look}.

As shown in Figs.~\ref{wrt-size}(a) and (b), the magnitude of the error in E$_b$ decreases with increasing cluster size in all XC functionals.
PBE@FLOSIC and SCAN@FLOSIC strongly overbind (by more than $\sim$150 meV/\ce{H2O}) the two small clusters, \ce{OH^-(H2O)} and \ce{H3O^+(H2O)}, and the errors tend to decrease in the larger clusters.
PBE@FLOSIC performs better than SCAN@FLOSIC in the larger clusters, particularly in deprotonated clusters (Fig.~\ref{wrt-size}(a)).
We find that self-interaction corrections weaken the strength of HBs and result in more accurate binding energies than found with the uncorrected functionals.
However, the accuracy of the self-interaction corrected functionals are inconsistent under variation of cluster size due to irregularities in the magnitude of the SIC to the binding of the clusters (E$_b^{\rm SIC}$).
The value of E$_b^{\rm SIC}$ is the difference between the FLOSIC--DFA and DFA@FLOSIC binding energies. 
As shown in Figs.~\ref{wrt-size}(c) and (d), E$_b^{\rm SIC}$ from FLOSIC--PBE is $\sim$20 meV/\ce{H2O} for the smallest as well as the largest deprotonated clusters, and it fluctuates in the range of 4--28 meV/\ce{H2O} for the intermediate-sized clusters, without showing any clear trend with respect to cluster size (Fig.~\ref{wrt-size}(c)).
With FLOSIC--SCAN, E$_b^{\rm SIC}$ is exceptionally small (4 meV/\ce{H2O}) in OH$^-$(H$_2$O).
This is because the total SIC computed from the OH$^-$(H$_2$O) molecule largely cancels the sum of the SIC energies of the isolated \ce{H2O} and OH$^-$ molecules.
In contrast, E$_b^{\rm SIC}$ is much larger in H$_3$O$^+$(H$_2$O), \textit{i.e.}, 109 meV/\ce{H2O} and 52 meV/\ce{H2O} with FLOSIC--PBE and FLOSIC--SCAN, respectively (Fig.~\ref{wrt-size}(d)).
E$_b^{\rm SIC}$ is reduced in both FLOSIC--PBE (19 meV/\ce{H2O}) and FLOSIC--SCAN (30 meV/\ce{H2O}) for the largest protonated cluster. 
These values are also similar to the E$_b^{\rm SIC}$ obtained in OH$^-$(H$_2$O)$_6$.
It is surprising that FLOSIC--SCAN provides too small E$_b^{\rm SIC}$ in OH$^-$(H$_2$O), as opposed to that in H$_3$O$^+$(H$_2$O). We explore this issue in more detail.

\begin{table}[htb]
 \caption{FLOSIC--SCAN self-interaction correction to the SCAN@FLOSIC total energy, defined as the correction from Eq.~(\ref{EPZ}) to the SCAN total energy evaluated on FLOSIC--SCAN Fermi-Löwdin orbitals, and its contributions from core, bond-pair, and lone-pair electrons. The pair contributions are written in the form $n\times \epsilon$, where $n$ is the number of pairs and $\epsilon$ is the average SIC energy of a pair.}
 \begin{ruledtabular}
  \begin{tabular}{ccccc}
    &  \multicolumn{4}{c}{Energy (meV)}\\
    \cline{2-5}
    System & Total & Core & Bond- & Lone- \\
           &       &      & pair  & pair \\
   \hline
   \ce{OH^-}          & 3342 & 1$\times251$ &  1$\times600$ & 3$\times830$ \\
   \ce{H2O}           & 2790 & 1$\times242$ & 2$\times545$ & 2$\times729$ \\
   \ce{H3O^+}         & 2599 & 1$\times277$ & 3$\times535$ &  1$\times717$ \\
   \hline
   \ce{OH^-(H2O)}     & 6136 & 2$\times245$ & 3$\times604$ & 5$\times767$ \\
   \ce{H3O^+(H2O)}    & 5441 & 2$\times251$ & 6$\times576$ & 2$\times742$ \\
  \end{tabular}
  \end{ruledtabular}
  \label{table-sic}
\end{table}

We note that the SIC energy obtained from each localized orbital (as given by the second term in Eq. (1)) with FLOSIC--SCAN is positive in these molecules, since for a given noded orbital density the SCAN XC energy is too negative compared to the exact XC energy (the negative of the Hartree energy). Since the SCAN total 
energy is already very accurate, this self-interaction ``correction" from noded orbital densities actually worsens it \cite{Shahi2019}.
\textcolor{black}{The} SIC energy in each lone-pair orbital is greater than that in each bond-pair orbital by $\sim$35\% in the isolated \ce{OH^-}, \ce{H2O}, and \ce{H3O^+} molecules.
Table~\ref{table-sic} shows the decomposition of the total SIC energy into contributions from core, lone-pair, and bond-pair orbitals. The self-interaction correction to the binding energy, E$_b^{\rm SIC}$, is only 6136-3342-2790 = 4 meV for OH$^-$(H$_2$O), too small to provide a significant correction to the SCAN@FLOSIC overbinding, but it is 5441-2599-2790 = 52 meV for H$_3$O$^+$(H$_2$O), where it provides a much more significant correction. This happens despite the lowering of the energy of H$_3$O$^+$(H$_2$O) due to the transfer of a pair of electrons from lone-pair to bond-pair FLOs when the strong hydrogen bond forms in that cationic cluster. 
%
% The normalized total SIC energy predominantly follows the relative fraction of lone-pair orbitals.
%
%As seen in Table~\ref{table-sic}, the total SIC in OH$^-$ is larger than others since six out of ten orbitals are lone-pairs.
%
%The decreasing fraction of lone-pair orbitals in \ce{H2O} (four out of ten) and H$_3$O$^+$ (two out of ten) reduces the total SIC per orbital.
%As computed with FLOSIC--SCAN the SIC energy for each lone-pair orbital is largest in OH$^-$ (411 meV) than in \ce{H2O} (367 meV) and H$_3$O$^+$ (362 meV).
%
%As a result, the total SIC energies computed with FLOSIC--SCAN in OH$^-$, \ce{H2O}, and H$_3$O$^+$ are respectively 3.342 eV, 2.790 eV, and 2.599 eV.
%
% In OH$^-$(H$_2$O), the bond-pair (lone-pair) SIC energy is in excess (deficit) compared to that obtained in OH$^-$ and \ce{H2O} combined.
% %
% The cancellation of these two quantities lead to the small SIC to the binding of OH$^-$(H$_2$O) with FLOSIC--SCAN.
%
%It turns out that the excess bond-pair SIC energy in OH$^-$(H$_2$O) compared to that from OH$^-$ and \ce{H2O} combined cancels with the deficit in the lone-pair SIC energy, leading to the small SIC to the binding of this molecule.
%
%As turned out, the FLOSIC--SCAN computed total SIC energy of OH$^-$(H$_2$O) largely cancels with the sum of the SIC energies in OH$^-$ and \ce{H2O}.
%
Thus the FLOSIC--SCAN method still lacks the balance required to treat both OH$^-$ and OH$^-$(H$_2$O) accurately enough to capture SIC as needed.
The description of H$_3$O$^+$ and H$_3$O$^+$(H$_2$O) with FLOSIC is much better in this regard, leading to a significant improvement in the binding of H$_3$O$^+$(H$_2$O).
\begin{figure}[htb]
\includegraphics[width=1.0\linewidth]{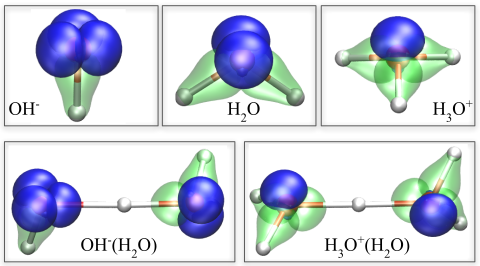}
\caption{Isosurfaces (0.02 $e$/\AA$^3$) of FLO densities from FLOSIC--SCAN for \ce{OH^-}, \ce{H2O}, \ce{H3O^+}, \ce{OH^-(H2O)}, and \ce{H3O^+(H2O)}, showing lone-orbitals in blue and bond-orbitals in green.}
\label{isosurface}
\end{figure}
\begin{figure}[htb]
%\centering  
% \includegraphics[width=1.0\linewidth]{halide-error-nolda.pdf}
\includegraphics[width=1.0\linewidth]{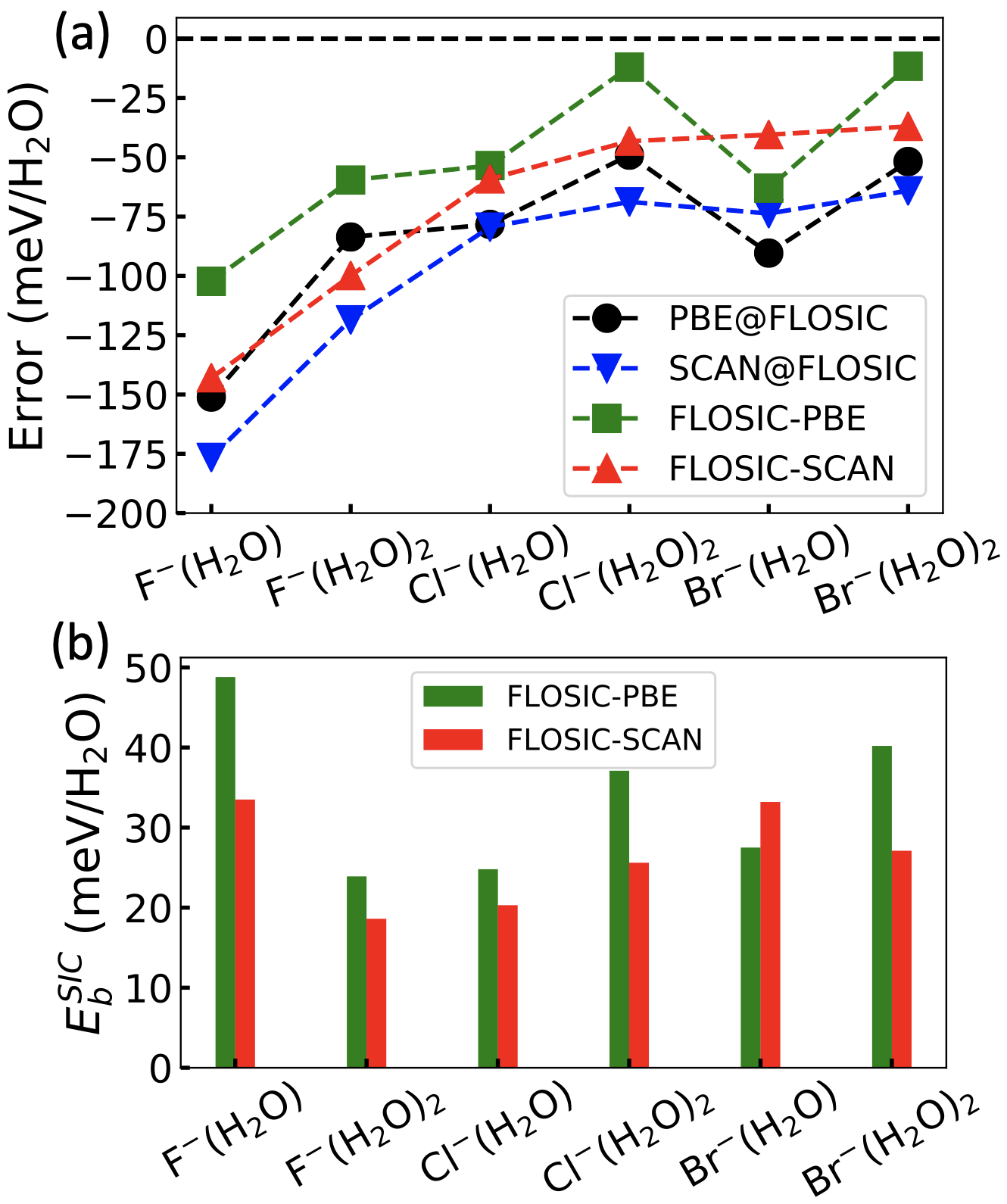}
\caption{(a) Error in the binding energies and (b) self-interaction correction to the binding energies ($E_b^{\rm SIC}$) of X$^{-}$(H$_{2}$O)$_{n}$ with X=(F, Cl, Br) and $n$=1,2 with various functionals compared to the CCSD(T)-F12b reference.}
\label{halide-water}
\end{figure}
\begin{table*}[htb]
 \caption{Binding energies with density functionals and the corresponding mean unsigned errors (MUEs) computed with respect to CCSD(T)-F12b reference~\cite{arismendi2016ttm} for halide-water clusters. The final column shows the percent error (PE) for FLOSIC--PBE. The energies are obtained with the basis DFO$^\ast$ for LSDA, PBE, and SCAN, and with \textit{diffuse}-3 for all other functionals.}
%  \centering
\begin{ruledtabular}
  \begin{tabular}{lccccccccccc}
    &     \multicolumn{10}{c}{Binding Energy per H$_2$O (meV/H$_2$O)}& PE\\
    \cline{2-11}
    Cluster & Ref. & LSDA & LSDA@  & FLOSIC-- & PBE & PBE@   & FLOSIC-- & SCAN & SCAN@  & FLOSIC-- & FLOSIC-- \\
            &      &      & FLOSIC & LSDA     &     & FLOSIC & PBE      &      & FLOSIC & SCAN     & PBE      \\
%    \cmidrule(rl){1-1}\cmidrule(rl){2-2}\cmidrule(rl){3-3}\cmidrule(rl){4-4}\cmidrule(rl){5-5}\cmidrule(rl){6-6}\cmidrule(rl){7-7}\cmidrule(rl){8-8}\cmidrule(rl){9-9}\cmidrule(rl){10-10}\cmidrule(rl){11-11}\cmidrule(l){12-12}
    \hline
     \ce{F^-(H_2O)} & -1188&	-1647&	-1638&	-1484&	-1350	&-1339&	-1290	&-1366&	-1364&	-1331&8.6 \%
 \\
     \ce{Cl^-(H_2O)} &-645&	-900&	-918&	-802&	-730	&-724	&-699	&-720&	-725&	-704&8.3 \%
 \\
     \ce{Br^-(H_2O)} &-556&	-785&	-811&	-669&	-634&	-646&	-619&	-624&	-630&	-596&11.3 \%
 \\
     \ce{F^-(H_2O)2} &-1056&	-1398&	-1394	&-1294&	-1142&	-1140&	-1116	&-1174&	-1175&	-1156&5.6 \%
 \\
     \ce{Cl^-(H_2O)2} &-648&	-901&	-903&	-804&	-699&	-697&	-660&	-717&	-717&	-691&1.9 \%
 \\
     \ce{Br^-(H_2O)2} &-575&	-814&	-816&	-717&	-625&	-626&	-586&	-640&	-639&	-612&2.0 \%
 \\
    \hline
    MUE   &--&	296&	302&	184&	85&	84&	50&	96&	97	&70&--
 \\
  \end{tabular}
\end{ruledtabular}  
\label{table-halide}
\end{table*}

\subsection{Water--Halide Ion Interactions}
In this section, we explore the interaction energy of water--halide dimers and trimers, each cluster containing one of the three halide anions, \ce{F^-}, \ce{Cl^-}, and \ce{Br^-}. 
As shown in Fig.~\ref{halide-str}(a), when the halide anion interacts with one \ce{H2O} molecule, the global minimum structures contain one hydrogen bond, in which a halide anion accepts HBs from a donor hydrogen.
With increasing size of the anion, the length of the HB increases from 1.39 \AA\ in \ce{F^-(H2O)} to 2.37 \AA\ in \ce{Br^-(H2O)}, and the HB angle moves away from linearity, indicating a systematic weakening of HBs with the size of anions.
%
%In the water--halide trimers each water molecule form one HB with the halide anion, and a similar elongation of the HBs is also found with increasing size of the anion. 

Table~\ref{table-halide} shows the binding energies of halide--water clusters and the mean unsigned errors (MUEs) obtained with XC functionals in comparison to CCSD(T)/F12b reference energies~\cite{egan2020nature}.
The reference values show that the magnitudes of the binding energies decrease by $\sim$50\% from lighter to heavier halide ions. 
This trend is qualitatively reproduced by all XC functionals, , and follows from the increasing ionic radii from F$^-$ to Br$^-$.
% , as well as that the the binding per molecule can decrease with an increasing number of \ce{H2O} molecules in the cluster.
% %
% Both of these trends are qualitatively reproduced by all XC functionals.
%
Here, too, FLOSIC--DFA methods systematically reduce the error in binding energies compared to the corresponding parent DFAs.
LSDA@FLOSIC predicts too large MUE (302 meV/\ce{H2O}) and FLOSIC--LSDA largely reduces the errors but still displays a large MUE of 184 meV/\ce{H2O}.
Both PBE@FLOSIC and SCAN@FLOSIC bring significant improvements compared to LSDA by bringing the MUE below 100 meV/\ce{H2O}.
FLOSIC--SCAN predicts an MUE of 70 meV/\ce{H2O}, reducing it by 27 meV/\ce{H2O} from SCAN@FLOSIC.
Again, FLOSIC--PBE provides the best agreement with the reference binding energies, predicting on average 50 meV/\ce{H2O} too-strong binding, after reducing the error by 34 meV/\ce{H2O} from PBE@FLOSIC.

Figure~\ref{halide-water}(a) further illustrates that the error in binding energies decreases with cluster size in all functionals.
The SIC contribution to the binding energy is in the range of 25--50 meV/\ce{H2O} with FLOSIC--PBE and 20--35 meV/\ce{H2O} with FLOSIC--SCAN, as shown in Fig.~\ref{halide-water}(b).
Except for \ce{Br^-(H2O)}, FLOSIC--PBE performs better than FLOSIC--SCAN for all water--halide clusters.

\begin{table*}[htb]
 \caption{Binding energies with density functionals and the corresponding mean unsigned errors (MUEs) computed with respect to CCSD(T)-F12b reference~\cite{arismendi2016ttm} for alkali-water clusters. The final column shows the percent error (PE) for FLOSIC--PBE. The energies are obtained with the basis DFO$^\ast$ for all functionals.
 }
%  \centering
\begin{ruledtabular}
  \begin{tabular}{@{}lccccccccccc@{}}
    &     \multicolumn{10}{c}{Binding Energy (meV/H$_2$O)} & PE\\
    \cline{2-11}
    Cluster & Ref. & LSDA & LSDA@  & FLOSIC-- & PBE & PBE@   & FLOSIC-- & SCAN & SCAN@  & FLOSIC-- & FLOSIC-- \\
            &      &      & FLOSIC & LSDA     &     & FLOSIC & PBE      &      & FLOSIC & SCAN     & PBE      \\
\hline
     \ce{Li^{+}(H_{2}O)} & -1508 &	-1619 &	-1651	& -1707 &	-1502	& -1531 &	-1596	& -1472 &	-1488 &	-1557&5.8 \%
 \\
     \ce{Na^{+}(H_{2}O)} &-1049&	-1160&	-1185 &	-1213&	-1035 &	-1048&	-1097&	-1062&	-1072 & -1127&4.5 \%
 \\
     \ce{K^{+}(H_{2}O)} &-779&	-867&	-887&	-906&	-731&	-749&	-794&	-762&	-771&	-823&1.9 \%
 \\
     \ce{Li^{+}(H_{2}O)_{2}} &-1404&	-1500&	-1527	&-1581&	-1392&	-1418&	-1481&	-1371&	-1384&	-1443&5.5 \%
 \\
     \ce{Na^{+}(H_{2}O)_{2}} &-993&	-1095&	-1115&	-1146&	-978&	-992&	-1043&	-1003	&-1012&	-1065&5.0 \%
 \\
     \ce{K^{+}(H_{2}O)_{2}} &-734&	-814&	-824&	-834&	-689&	-704&	-729&	-718&	-726&	-761&-0.7 \%
 \\
    \hline
    MUE   &--&	98&	120&	153&	23&	17&	47&	21&	16&	51&--
 \\
   \hline
  \end{tabular}
\end{ruledtabular}  
\label{table-alkali}
\end{table*} 

\subsection{Water--Alkali Ion Interactions}

In this section, we explore the interaction energy of water--alkali cation dimers and trimers, each cluster containing one of the three alkali cations, \ce{Li+}, \ce{Na+}, and \ce{K+}. 
The nature of interactions in the global minimum structures of the small water--alkali clusters is different from all the other clusters discussed so far.
The water--alkali interaction is not due to HBs, but ion--dipole interactions, in which a cation sits close to a more electronegative oxygen atom, as shown in Fig.~\ref{halide-str}(b).
With the increasing size of the cation, the oxygen-cation distance increases from 1.84 \AA\ in \ce{Li^+(H2O)} to 2.60 \AA\ in \ce{K^+(H2O)}, indicating weakening of the binding energy.
Table~\ref{table-alkali} shows the binding energies of alkali--water clusters and the MUEs obtained with XC functionals in comparison to the CCSD(T)/F12b reference~\cite{arismendi2016ttm}.
The reference binding energies show that the binding energies are reduced by $\sim$50\% from the lighter (\ce{Li^+}) to the heavier (\ce{K^+}) alkali ion, and the binding per molecule decreases with an increasing number of \ce{H2O} molecules in the cluster.
Both of these trends are qualitatively reproduced by all XC functionals.
Unlike the results found in the hydrogen-bonded clusters, the FLOSIC--DFA functionals do not improve the binding energies of alkali--water clusters compared to the parent DFAs.
The MUEs from FLOSIC--DFA functionals are 30--35 meV/\ce{H2O} worse than the corresponding DFAs.
Both PBE@FLOSIC and SCAN@FLOSIC predict $\sim$16 meV/\ce{H2O} MUE, which is the lowest among all functionals.

\begin{figure}[htb]
\includegraphics[width=0.99\linewidth]{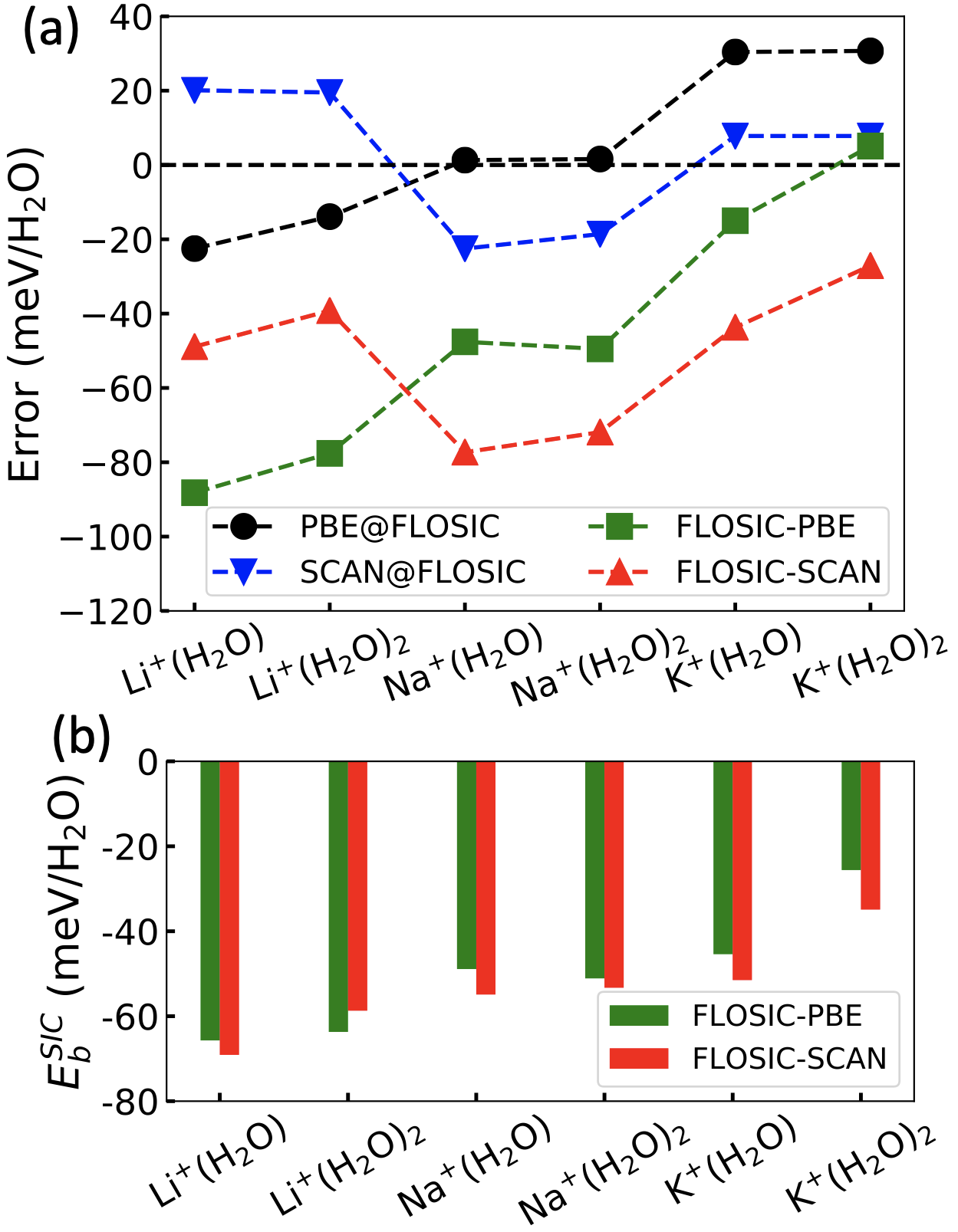}
\caption{(a) Error in the binding energies and (b) self-interaction correction to the binding energies ($E_b^{\rm SIC}$) of M$^{+}$(H$_{2}$O)$_n$ (M=Li, Na, K) and $n$=1,2 with various functionals compared to the CCSD(T)-F12b.}
\label{alkali-water}
\end{figure}

Fig.~\ref{alkali-water} (a) shows the error in binding energies for different clusters.
With PBE@FLOSIC, the magnitudes of the binding energies are overestimated in the lighter \ce{Li^+(H2O)} cluster and underestimated in the heavier cation \ce{K^+(H2O)} cluster.
In the case of SCAN@FLOSIC, the errors tend to decrease from the lighter to the heavier cation cluster.
Using FLOSIC--PBE and FLOSIC--SCAN, the average $E_b^{\rm SIC}$ is $\sim$50 meV/\ce{H2O}, with its magnitude declining towards heavier water--alkali clusters, as shown in Fig.~\ref{alkali-water}(b).

For the water--alkali clusters, unlike the other clusters studied here, self-interaction errors seem to be unimportant.

\begin{figure}[htb]
\centering  
\includegraphics[clip, width=0.99\linewidth]{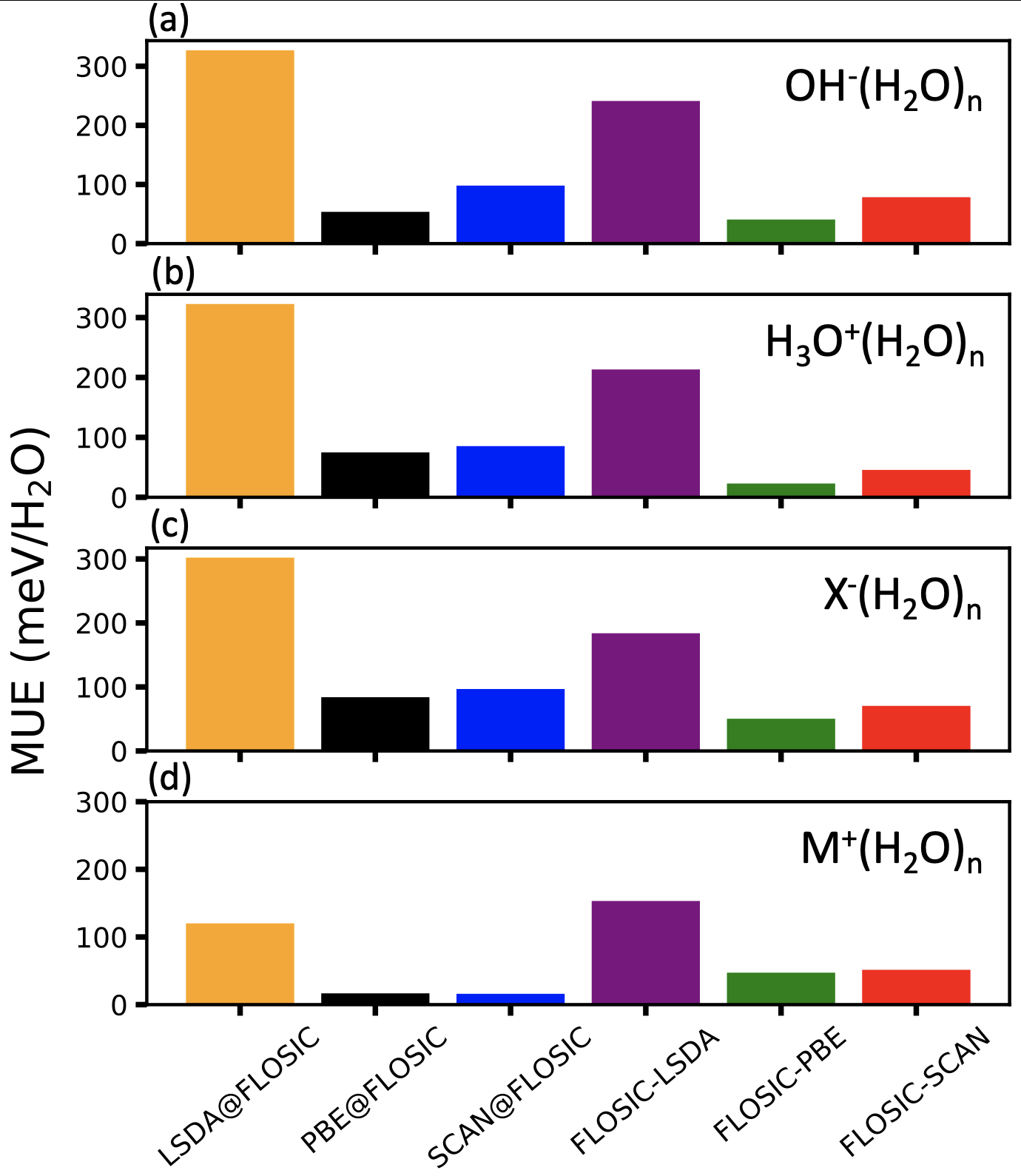}
\caption{Mean unsigned error (MUE) given by different functionals for (a) deprotonated water, (b) protonated water, (c) water--halide, and (d) water--alkali clusters, shown from the top to the bottom.}
\label{mue-overall}
\end{figure}

\section{Conclusions}
We have assessed the accuracy of the orbital-by-orbital PZ self-interaction correction~\cite{PerZun-PRB-81} computed within the FLOSIC methodology~\cite{YanPedPer-PRA-17} for the interaction energies between water molecules and various ions, namely, hydroxide, hydronium, halide anions, and alkali cations in gas-phase clusters.
These clusters are primarily hydrogen-bonded, except the water--alkali clusters which are bound by ion--dipole interactions.
It is known that negatively-charged clusters are subject to erroneous delocalization of the electron density due to self-interaction present in approximate XC functionals such as local and semi-local DFAs. 
We have employed FLOSIC--DFA methods, where the DFAs are LSDA, PBE, and SCAN, to mitigate this error in the electron density and in the corresponding energetics.
The self-interaction corrected density is utilized to quantify the density-driven error in the energies obtained with these DFAs.
We have found that the magnitude of the density-driven error increases with diffuse Gaussian functions, which makes the energy convergence of anions with respect to basis functions more ambiguous.
The density-driven error is not significant in cations.
The self-interaction corrected density allows a meaningful basis-set convergence study for anions with LSDA, PBE, and SCAN.
The DFA energies computed here with the \textit{diffuse}-3 basis at the FLOSIC--DFA densities, \textit{i.e.}, DFA@FLOSIC, provide an accurate assessment of the accuracy of these functionals for anions.

Fig.~\ref{mue-overall} shows the summary and cross-comparison of the accuracy of the methods for the four types of water--ion clusters.
The DFA@FLOSIC energies show a significant overestimation of the binding energies for the hydrogen-bonded clusters, \textit{i.e.}, protonated water, deprotonated water, and halide--water clusters.
The MUE is very large in LSDA@FLOSIC ($>$300 meV/\ce{H2O}) and is significantly reduced in PBE@FLOSIC (54--84 meV/\ce{H2O}) and SCAN@FLOSIC (85--98 meV/\ce{H2O}).
The removal of SIE weakens the binding in hydrogen-bonded clusters and shifts the binding energies toward the accurate references.
Surprisingly, the average reduction of the SIE is larger in the binding of protonated water clusters than in deprotonated and water--halide clusters.
Because of that, all FLOSIC--DFA functionals produce much smaller MUE for the protonated water clusters than for the other hydrogen-bonded clusters.
FLOSIC--SCAN predicts the MUE in the range of 46--78 meV/\ce{H2O}.
FLOSIC--PBE provides the most accurate description of the hydrogen-bonded clusters, with the MUE in the range of 23--47 meV/\ce{H2O}.
The effect of removing SIE in water--alkali clusters is in stark contrast to that in hydrogen-bonded clusters.
Self-interaction correction strengthens the binding of the alkali--water clusters, and as a result, the MUE from FLOSIC--DFAs are worse than the corresponding DFAs. This is in line with the fact that, in other situations, PZ-SIC reduces self-interaction errors but introduces other errors for many-electron densities \cite{Shahi2019}, as discussed in the next paragraph.
%
%On average, both PBE@FLOSIC and SCAN@FLOSIC provide reasonably accurate binding energies within $\sim$16 meV/\ce{H2O} of the reference for water--alkali clusters.

Although FLOSIC--DFAs improve the description of the hydrogen-bonded clusters, the overall results suggest that the PZ SIC method needs improvement in order to achieve more accurate binding energies, in particular, for the small hydrogen-bonded clusters and the alkali--water clusters.
The PZ SIC approach is exact for all one-electron densities, however, it is not so accurate in diverse many-electron regions~\cite{Vydrov2006, Santra2019}.
An appropriate scaling of the PZ SIC is still required to make it equally accurate in many-electron regions. 
A scaling method that would enhance the SIC contribution to the binding of \ce{OH^-(H2O)} and \ce{H3O^+(H2O)} clusters and simultaneously reduce the SIC contribution to the binding of water--alkali clusters is highly sought after.
\textcolor{black}{A few approaches that locally scale down the PZ SIC
have been introduced recently~\cite{Santra2019,zope2019,Bhattarai2020,Bhattaraienhancing}, but more development is needed.
The scaled methods have not yet been implemented self-consistently. When implemented on FLOSIC-LSDA FLO densities, the methods of Refs. \citenum{Santra2019} and \citenum{Bhattaraienhancing} improve many calculated properties over PZ SIC, including the energies of the stronger bonds~\cite{Santra2019,zope2019,Bhattarai2020,Bhattaraienhancing, Li2020},  but seriously underbind~\cite{Bhattaraienhancing} typical hydrogen-
% the methods of 
% Refs.~\cite{Santra2019,Bhattaraienhancing}, have been found~\cite{Bhattaraienhancing} to seriously underbind the hydrogen- 
%
and van der Waals-bonded complexes.
Our current study with self-consistent FLOSIC implementations of unscaled PZ SIC shows the importance of mitigating} the self-interaction error from the electron density and the corresponding energetics in water--ion interactions. We aim to increase the accuracy of self-interaction correction in future works.

% A few approaches of scaling PZ SIC have been introduced recently~\cite{zope2019,Bhattarai2020,Yamamoto2020}, but more developments are needed along this direction.
%
% Nevertheless, this study shows the importance of mitigating the self-interaction error from the electron density and the corresponding energetics in water--ion interactions. We aim to increase the accuracy of self-interaction correction in future works.    

%FLOSIC--PBE is giving the overall best accuracy among the density functional methods we studied. It improves the mean unsigned error (MUE) of deprotonated water clusters and protonated water clusters by 24.21\% and 69.43\% respectively, with respect to the underlying density corrected PBE. Performance of FLOSIC--SCAN is also significant but not as good as that of FLOSIC--PBE. FLOSIC--SCAN improved the MUE of deprotonated cluster by 24.21 \% and MUE of protonated cluster by 46.49\% relative to underlying density corrected SCAN functional. Although FLOSIC--LSDA improved significantly over LSDA(@FLOSIC), it is still overbinding the clusters very strongly. It has also been observed that magnitude of the error decreases as the cluster size increases, for both deprotonated and protonated water clusters.

\section*{Supplementary material}
The supplementary material includes the optimized FOD coordinates, NRLMOL inputs, and total energies with all methods for the purpose of data reproduction.

\begin{acknowledgments}
This work was supported by the U.S. Department of Energy, Office of Science, Office of Basic Energy Sciences under award number DE-SC0018331 as a part of the Computational Chemical Sciences Program. 
The work of K.W. and P.B was supported by the U.S. National Science Foundation under Grant No. DMR-1939528. Calculations for this work were done on Temple University's HPC resources and thus were supported in part by the National Science Foundation grant number 1625061 and by the US Army Research Laboratory under contract number W911NF-16-2-0189. K.W., B.S., and J.P.P acknowledge Francesco Paesani and Colin Egan for providing CCSD(T) reference binding energies for halide-water and alkali-water clusters.
\end{acknowledgments}

\section*{Data Availability}
The data that support the findings of this study are available within the article and its supplementary material and from the
authors upon reasonable request.

%\nocite{*}
%\bibliography{aipsamp}% Produces the bibliography via BibTeX.
\bibliography{reference}

\end{document}